\documentclass[aps,prl,reprint,superscriptaddress,floatfix,amssymb,amsfonts,showpacs]{revtex4-1}

\usepackage{graphicx}
\usepackage[version=3]{mhchem}
\usepackage{lipsum}
\usepackage{graphicx}
\usepackage{dcolumn}

\newcommand{\forceindent}{\leavevmode{\parindent=1em\indent}}

\begin{document}

\title{Strong Spin-Phonon Coupling Unveiled by Coherent Phonon Oscillations in Ca$_2$RuO$_4$}

\author{Min-Cheol Lee}
\author{Choong H. Kim}
\author{Inho Kwak}
\affiliation{Center for Correlated Electron Systems (CCES), Institute for Basic Science (IBS), Seoul 08826, Republic of Korea}
\affiliation{Department of Physics and Astronomy, Seoul National University, Seoul 08826, Republic of Korea}
\author{C. W. Seo}
\affiliation{Center for Correlated Electron Systems (CCES), Institute for Basic Science (IBS), Seoul 08826, Republic of Korea}
\affiliation{Department of Physics, Chungbuk National University, Cheongju, Chungbuk 28644, Republic of Korea}
\author{Changhee Sohn}
\affiliation{Center for Correlated Electron Systems (CCES), Institute for Basic Science (IBS), Seoul 08826, Republic of Korea}
\affiliation{Department of Physics and Astronomy, Seoul National University, Seoul 08826, Republic of Korea}
\author{F. Nakamura}
\affiliation{Department of Education and Creation Engineering, Kurume Institute of Technology, Fukuoka 830-0052, Japan}
\author{C. Sow}
\author{Y. Maeno}
\affiliation{Department of Physics, Graduate School of Science, Kyoto University, Kyoto 606-8502, Japan}
\author{E.-A. Kim}
\affiliation{Department of Physics, Cornell University, Ithaca, New York 14853, USA}
\author{T. W. Noh}
\thanks{Corresponding author}
\email{twnoh@snu.ac.kr}
\affiliation{Center for Correlated Electron Systems (CCES), Institute for Basic Science (IBS), Seoul 08826, Republic of Korea}
\affiliation{Department of Physics and Astronomy, Seoul National University, Seoul 08826, Republic of Korea}
\author{K. W. Kim}
\thanks{Corresponding author}
\email{kyungwan.kim@gmail.com}
\affiliation{Department of Physics, Chungbuk National University, Cheongju, Chungbuk 28644, Republic of Korea}

\date{\today}

\begin{abstract}
We utilize near-infrared femtosecond pulses to investigate coherent phonon oscillations of Ca$_2$RuO$_4$. The coherent $A_g$ phonon mode of the lowest frequency changes abruptly not only its amplitude but also the oscillation-$phase$ as the spin order develops. In addition, the phonon mode shows a redshift entering the magnetically ordered state, which indicates a spin-phonon coupling in the system. Density functional theory calculations reveal that the $A_g$ oscillations result in octahedral tilting distortions, which are exactly in sync with the lattice deformation driven by the magnetic ordering. We suggest that the structural distortions by the spin-phonon coupling can induce the unusual oscillation-$phase$ shift between impulsive and displacive type oscillations.
\end{abstract}

\maketitle

 Ultrashort pulses provide unique opportunities to study material properties in the time domain. The real-time observation on a femtosecond time scale after external stimuli allows us to single out important interactions during the recovery back to the equilibrium state \cite{Kim2012, Gerber2015, Fausti2014, Mankowsky2014, Huber2001, Brorson1990, Schmitt2008, Kubler2007, Schafer2010, Forst2011, Polli2007, Eschenlohr2013, Dhar1994, Stevens2002, Zeiger1992, Garrett1996, Li2013, Riffe2007}. Coherent oscillations are in the spot-light of recent studies of ultrafast phenomena not only in the scientific aspect but also for the ultrafast control of the material properties  \cite{Kim2012, Gerber2015, Schmitt2008, Kubler2007, Schafer2010, Dhar1994, Stevens2002, Forst2011, Polli2007, Zeiger1992, Garrett1996, Li2013, Riffe2007}. That is, the coherent lattice motions have been claimed to play the major role for the photoinduced high-temperature superconductivity in cuprates and also for the transient magnetic order in Fe-based superconductors \cite{Kim2012, Gerber2015, Fausti2014, Mankowsky2014}. Therefore, it is important to understand and control the coherent oscillations.\\
 \begin{figure}[b!]
	\centering
	\includegraphics[width = 3.5 in]{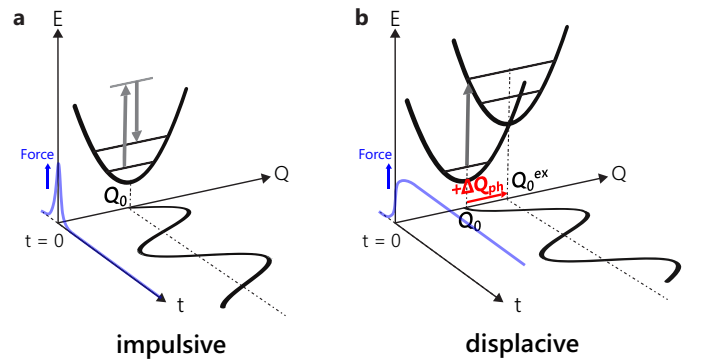}
	\caption{(color online) Generation mechanisms of coherent oscillations. Whereas the impulsive stimulated Raman scattering does not alter the average lattice coordinates, a displacive excitation can cause a shift of the coordinates in the excited state (from \ce{$Q_0$} to \ce{$Q_0^{ex}$}). There should be light absorption to generate the displacive oscillations, while the impulsive ones require no absorption.}
	\label{Figure1}
\end{figure}
\forceindent The generation of coherent oscillations has been explained by two types of excitation mechanisms, which are schematically illustrated in Fig. 1. Vibrations with sine-type modulations can be induced by impulsive stimulated Raman scattering process, resulting in sine-type oscillations \cite{Dhar1994, Stevens2002}. It explains the coherent oscillations observed in transparent compounds under pumping with a photon energy smaller than an optical gap. Alternatively, in an opaque material, the abrupt modification of the electron density distribution due to the absorption of pump photons can trigger displacive motions of ions towards new coordinates in the excited state, resulting in cosine-type oscillations \cite{Stevens2002, Zeiger1992, Garrett1996, Li2013, Riffe2007, Franck1926}. In principle, both mechanisms can be active in a single material \cite{Stevens2002, Garrett1996}. However, once the displacive mechanism has become active in an opaque material, it dominates the oscillation behavior \cite{Stevens2002, Franck1926}. Therefore, the oscillation-$phase$ of coherent phonons has been assumed to vary negligibly as long as the absorption of pump pulses does not significantly change \cite{Stevens2002, Zeiger1992, Garrett1996, Li2013}. To our surprise, however, we find that the coherent oscillations in Ca$_2$RuO$_4$ are exempted from these theories.
\\
\forceindent Ca$_2$RuO$_4$, a prototype 4$d$ Mott insulator, shows interesting phase transitions due to coupling among quantum degrees of freedoms \cite{Braden1998, Friedt2001, Jain2017, Das2018, Souliou2017}. Recent interest has been focused on the Van Vleck-type antiferromagnetic order below $T_{\textnormal{N}}$ = 113 K due to the sizable spin-orbit coupling \cite{Jain2017, Das2018, Souliou2017, Khaliullin2013}. In particular, collective Higgs oscillations of the magnetic moment have been discerned in an $A_g$ channel Raman spectrum, where involvement of phonons was suggested as a potential origin of the magnetic excitation in the fully symmetric Raman configuration \cite{Jain2017, Souliou2017}. However, the spin-phonon coupling has not been well identified yet in Ca$_2$RuO$_4$.

 In this paper, we demonstrate that the spin-phonon coupling controls the birth of coherent phonon oscillations in Ca$_2$RuO$_4$. We investigate coherent lattice vibrations of Ca$_2$RuO$_4$ that show enormous anomalies in the $A_g$ phonon mode of the lowest frequency across $T_{\textnormal{N}}$. Surprisingly, the anomalies include a huge oscillation-$phase$ shift by 90 degrees, which implies a change of the generation mechanisms of the coherent phonon due to the magnetic order. We find that octahedral distortions across $T_{\textnormal{N}}$ are in sync with the $A_g$ phonon mode from Density functional theory (DFT) calculations and suggest that the spin-phonon coupling is responsible for the unusual oscillation-\textit{phase} shift.

 We perform photoinduced reflectivity change measurements on single crystals of Ca$_2$RuO$_4$ \cite{Brorson1990, Schafer2010}. Samples are grown by the floating zone method \cite{Nakatsuji2001}. We use near infrared pulses generated from a commercial Ti:sapphire amplifier system with a 250 kHz repetition rate. The center wavelength of the pulses is about 800 nm, whose photon energy ($\sim$ 1.55 eV) is much larger than the optical gap ($\sim$ 0.3 eV) of Ca$_{2}$RuO$_{4}$ \cite{Jung2003}. The time duration of pump and probe pulses are 30 fs. The pump and probe pulses are linearly polarized and perpendicular to each other. The sample does not exhibit anisotropy depending on both pump and probe polarizations.

\begin{figure}[!t]
	\includegraphics[width=3.5 in]{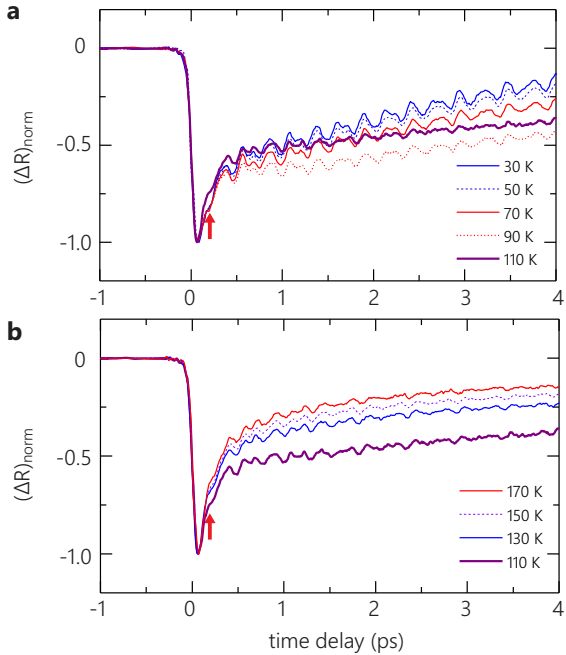}
	\caption{(color online) Temperature dependent photoinduced reflectivity changes normalized by maximum values ($\Delta R$)$_{\textrm{norm}}$. Photoinduced reflectivity is measured at every 20 K (a) from 30 K to 110 K ($T \lesssim T_\textnormal{N}$) and (b) from 110 K to 170 K ($T \gtrsim T_\textnormal{N}$)}
	\label{Figure2}
\end{figure}
 Figure 2 shows the temperature($T$)-dependent reflectivity changes normalized by maximum values ($\Delta R$)$_{\textrm{norm}}$ from 30 K to 170 K. The overall relaxation dynamics gradually changes across $T_{\textnormal{N}}$. In addition, periodic reflectivity modulations due to coherent lattice oscillations are clearly observed at all temperatures. A careful look on the oscillations in Fig. 2 can already capture an anomaly across $T_{\textnormal{N}}$ as indicated by the red arrow. That is, while the oscillation patterns are well aligned for $T > T_\textnormal{N}$, and for $T < T_\textnormal{N}$, respectively, there is an abrupt change across $T_{\textnormal{N}}$. 

\begin{figure*}[]
	\includegraphics[width = 7 in]{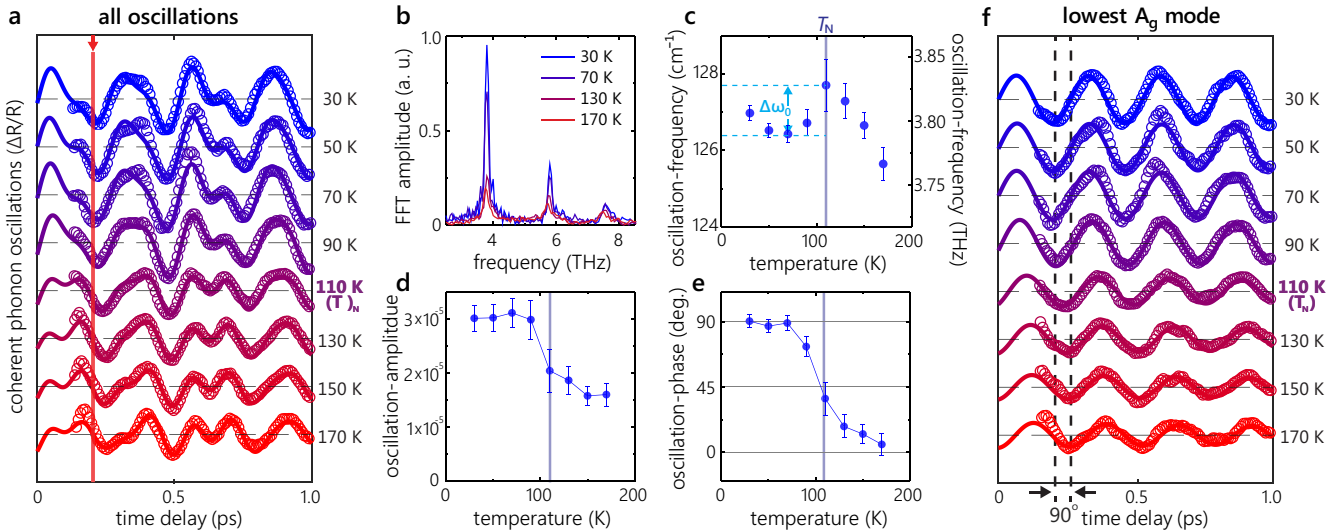}
	\caption{(color online) The (a) coherent phonon oscillations of photoinduced reflectivity after subtracting the electronic response by means of bi-exponential decay fits at temperatures of every 20 K from 30 K to 170 K. The red arrow in (a) indicates the same delay time marked by an arrow in Fig. 2. (b) Fourier-transform of the oscillations in (a). All the oscillating components correspond to the $A_g$ symmetric Raman modes. (c-e) $T$-dependent fitting parameters of the frequency, amplitude, and $phase$ of the lowest-frequency mode extracted from damped harmonic oscillator model fits. (f) The oscillating component and fit curves for the lowest-frequency mode. Higher-frequency oscillations have been subtracted by using the model fits. Dotted vertical lines contrast the oscillation-$phase$s at different temperatures.}
	\label{Figure3}
\end{figure*}
 Here, we focus on the coherent phonon oscillations. After subtracting the overall electronic responses, we plot the oscillation components at various temperatures in Fig. 3(a). Their Fourier transform, displayed in Fig. 3(b), reveals that multiple phonon modes are excited by the pump pulses. All of the modes correspond to the $A_g$ symmetric  phonon modes reported in previous Raman measurements \cite{Rho2005, Souliou2017}. To obtain further insight, we analyze the coherent phonons using a damped harmonic oscillator model: $\Delta{R}_{CP}(t)$= $-\Sigma_i{A_i}\cos(2\pi{f}_i+\phi_i)\exp(t/\tau_i)$, where ${A_i}, {f_i}, {\phi_i},$ and ${\tau_i}$ are the amplitude, frequency, initial $phase$, and damping time of the $A_g$ phonon modes. Fitting results are plotted as line curves in Fig. 3(a), which are well-matched to the measured oscillations.

 We find that the coherent phonon oscillations of the lowest frequency $A_g$ mode present anomalies across the magnetic ordering. Figures 3(c)-(e) show the $T$-dependence of the fitting parameters of the $A_g$ mode. All the parameters show clear anomalies across $T_{\textrm{N}}$. The oscillation component of the mode in Fig. 3(f) shows that the overall oscillations are dominated by this mode and the anomaly noted in Fig. 2 is due to the oscillation-$phase$ shift of the mode across $T_{\textrm{N}}$. The abrupt redshift of the phonon frequency with the magnetic ordering shown in Fig. 3(c) is a clear evidence of a strong spin-phonon coupling expected in Ca$_{2}$RuO$_{4}$ \cite{Sohn2017}. We point out that the subtraction of the electronic response is not the origin of the anomalies (see Fig. S2) as already clear in the raw data in Fig. 2, and confirm that the anomalies in the lowest mode do not depend on the fitting parameters of the higher frequency modes (see section 4 of the Supplemental Material \cite{SM}).

 The most remarkable is the anomaly in $\phi_{i}$, \textit{i.e.} the oscillation-$phase$ shift by 90 degrees across $T_{\textrm{N}}$. The change from cosine-type to sine-type oscillations implies that the generation mechanism changes from the displacive one to the impulsive one across the magnetic transition. As discussed in Fig. 1, both the impulsive and displacive characters of coherent oscillations can coexist, resulting in a random $phase$ such that $\Delta Q(t) = Asin(\Omega t) + Bcos(\Omega t) = \sqrt{A^{2}+B^{2}}cos(\Omega t + \phi)$ with $\phi = -arctan(A/B)$. According to the microscopic model of the coherent phonon generation by Stevens $\textit{et al.}$, the $phase$ is determined by the complex dielectric function $\tilde{\varepsilon} = \varepsilon_{1}+i\varepsilon_{2}$ \cite{Stevens2002, Li2013, Riffe2007}. When $\tilde{\varepsilon}$ varies slowly within the bandwidth of the pump pulse and the pulse duration is much shorter than the phonon oscillation period, the two-band approximation of the Raman susceptibility tensor results in $|A| \propto \frac{d\varepsilon_{1}}{d\omega}$ and $|B| \propto \frac{2\varepsilon_{2}}{\Omega}$ such that $|\phi| = arctan(\dfrac{d\varepsilon_{1}/d\omega}{2\varepsilon_{2}/\Omega})$ \cite{Stevens2002, Li2013, Riffe2007, SM}. From an ellipsometry measurements on Ca$_{2}$RuO$_{4}$, we find that $\frac{2\varepsilon_{2}}{\Omega} \sim$ 500 eV$^{-1}$ and $\frac{d\varepsilon_{1}}{d\omega}\lesssim$ 2 eV$^{-1}$ (see Fig. S6), which expects displacive oscillations of $\phi \approx 0$, at all measured temperatures across the magnetic phase transition. We note that if the displacive force due to pump photon absorption decays out fast, the $|\phi|$ can increase \cite{Li2013, Riffe2007}. If the force should relax much faster in the AFM state, then the oscillation amplitude must become smaller than that in the purely displacive case. Apart from that the theory does not suggest a sudden relaxation of the force in the AFM state, however, the observed amplitude is even larger in the AFM state on the contrary to this expectation. Therefore, the abrupt changes in the oscillation-$phase$ and amplitude shown in Fig. 3 are out of the scope of currently available theories. \\
\begingroup
\begin{table}[!b]
	\begin{ruledtabular}
		\begin{tabular}{l c c c}
			$phase$ & $experiment$ & $calculation$ \\
			\hline AFM phase & 126.4 cm\ce{^{-1}} (110 K) & 122.8 cm\ce{^{-1}} \\
			PM phases & 127.7 cm\ce{^{-1}} (70 K) & 125.4 cm\ce{^{-1}} \\
			difference (\ce{$\Delta\omega$_0})& -1.3 cm\ce{^{-1}} & -2.6 cm\ce{^{-1}} \\
		\end{tabular}
	\end{ruledtabular}
	\caption{Comparison of the lowest $A_g$ phonon frequencies in the antiferromagnetic (AFM) and paramagentic (PM) phases from our experiments and DFT calculations.}
	\label{Table1}
\end{table}
\endgroup
 Why should this mode respond sensitively to the spin texture? The abrupt changes of the phonon dynamics across the magnetic ordering temperature suggest the strong spin-phonon coupling in the system should play a role. We investigate the spin-phonon coupling by DFT calculations. The phonon eigenmode is shown in Fig. 4(a). We find that the phonon mode is dominantly of a tilting character. To evaluate the redshift in the magnetic state, we calculate the stiffness constant, which are proportional to the square of the oscillation frequency. The stiffness constants under magnetic order can be modified by the magnetic interaction as below:
\begin{equation}\label{key}
\tilde{C}_{\mu\nu}^{AFM}=\tilde{C}_{\mu\nu}^{PM}+ \dfrac{\partial^2}{\partial x_{\mu}\partial x_{\nu}}{J<{\tilde{S_i}}\cdot{\tilde{S_j}}}>
,\end{equation}
where $\tilde{C}_{\mu\nu}^{AFM}$  ($\tilde{C}_{\mu\nu}^{PM}$) is elastic stiffness constants under antiferromagnetic (AFM) phase (paramagnetic (PM) phase). $x_{\mu}$, $x_{\nu}$ are strain components, $J$ is the exchange interaction and \ce{$\tilde{S_i}$} is the pseudospin of $i$-th Ru ions, where spin and orbital moments are entangled as $\tilde{S}=\vec{S}+\vec{L}$ \cite{Jain2017, Khaliullin2013}. We simulate the PM phase by averaging the antiferromagnetic and ferromagnetic (FM) phases \cite{Fang2004}. Table 1 shows the calculated phonon frequency depending on the spin configuration. Our DFT calculations can well explain the observed redshift due to the spin-phonon coupling.

 How can the spin-phonon coupling introduce the anomalies in the coherent oscillations? To obtain further insights, we calculate the crystal structures under various spin configurations but with the lattice constants fixed with the reported values at 11 K \cite{Braden1998}. Figure 4(b) shows the positions of apical ($O_{\textrm{A}}$) and in-plane oxygen ($O_{\textrm{P}}$) atoms obtained from the DFT calculations depending on spin configurations. Close scrutiny into the octahedral structures reveals that the octahedral distortion across the magnetic transition is in sync with the eigenmode of the $A_g$ phonon shown in Fig. 4(a). The red arrows in Fig. 4(b) indicate the octahedral distortions by the $A_g$ phonon with an amplitude +$\Delta Q_{ph}$ of 0.015(0.011){\AA} in $O_{\textrm{A}}$($O_{\textrm{P}}$) in the PM state. Note that the ends of the arrows almost coincide with the positions of those atoms in the AFM state expected by the calculations. In addition, the arrows correspond to a decrease of the tilting angle of $O_{\textrm{A}}$ by 0.15 degrees in the AFM state. We note that the anomaly in the $O_{\textrm{A}}$ tilting has indeed been observed experimentally although it has not been noticed before because the anomaly is comparable to the experimental errorbar (see Fig. S9) \cite{Braden1998}. Therefore, we argue that the AFM order can induce the octahedral lattice deformation along with the $A_g$ mode.

 We suggest that the lattice deformation along the $A_g$ phonon mode across the magnetic phase transition should be responsible for the unusual oscillation-$phase$ change. As shown in Fig. 1(b), the $A_g$ phonon exhibits the cosine-type oscillations at $T>T_\textnormal{N}$ due to a shift of the lattice coordinates in the excited state from $Q_0$ to $Q^{ex}_0$. However, the sine-type oscillations in the AFM state implies that the lattice coordinates do not change by photon pumping once the spins order. This discrepancy in the lattice coordinates could be compromised if the AFM order induces a shift of the lattice coordinates ($+\Delta Q_{ph}$) in the same way that the photon pumping does in the PM state (from $Q_0$ to $Q^{ex}_0$). As revealed by our DFT calculations, the magnetic order indeed results in a lattice deformation along the phonon eigenmode in the equilibrium state. Therefore, the displacive character of the coherent oscillations could be well suppressed in the AFM state. 

\begin{figure}[!t]
	\centering
	\includegraphics[width=3.5 in]{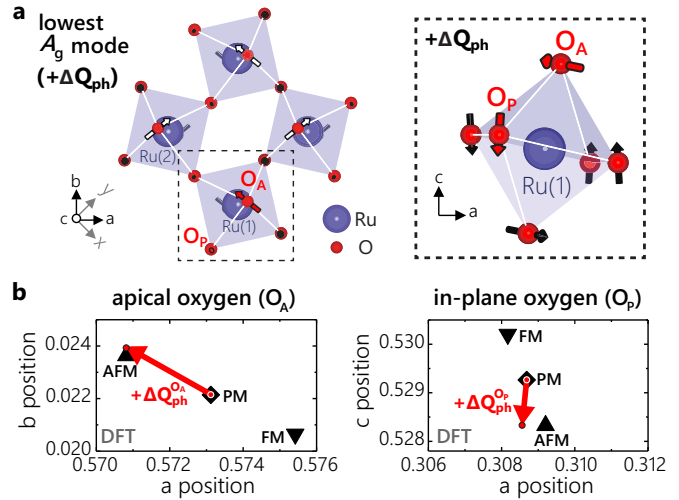}
	\caption{
	(color online) (a) Eigenmode of the lowest $A_g$ phonon derived by DFT calculations. The depicted vibration of the $A_g$ mode $+\Delta Q_{ph}$ corresponds to the direction represented by arrows in (b). (b) Positions of the apical ($O_{\textrm{A}}$) and in-plane oxygen ($O_{\textrm{P}}$) atoms noted in (a) expected from the DFT calculations under various spin configurations. All the position values are normalized by each unit cell lengths. The displacements of both oxygen atoms from the PM phase (black diamond) to the AFM phase (black triangle) are close to the motion of the lowest $A_g$ phonon with an amplitude of 0.015(0.011) \ce{\AA} in $O_{\textrm{A}}$($O_{\textrm{P}}$) ($+{\Delta}Q_{ph}$; red arrows).	
	}
	\label{Figure4}
\end{figure}
 Although the spin-phonon coupling can explain the abrupt shift of the oscillation-$phase$ qualitatively, the larger amplitude of the impulsive oscillations in the AFM state shown in Fig. 3(d) requires a further theoretical study. Within the two-band model of the Raman tensor, the oscillation-$phase$ is governed by the dielectric functions and the decay time of the displacive force, while the overall amplitude can be modified by the matrix element of the electron-phonon coupling \cite{Stevens2002, Riffe2007, Li2013}. The two-band approximation of the Raman tensor nicely explains coherent oscillations even in materials where more than two bands are involved \cite{Kim2012, Gerber2015}. However, it seems not working on Ca$_{2}$RuO$_{4}$.
  
 We propose that it is important to understand the lattice deforamtion driven by the magnetic ordering in the context of the Van-Vleck type magnetism in Ca$_{2}$RuO$_{4}$. The magnetic order in Ca$_{2}$RuO$_{4}$ is determined by the spin-orbital entangled pseudospin $\tilde{S}=\vec{S}+\vec{L}$ \cite{Jain2017, Das2018}. The order of the spin-orbit coupled states can naturally induce a change in the orbital configuration \cite{Das2018}, which can influence the crystal structure as revealed by our DFT calculations. We suggest that the strong spin-phonon coupling of Ca$_{2}$RuO$_{4}$ may alter the Raman tensor across a magnetic phase transition. We note that the absorption in Ca$_{2}$RuO$_{4}$ is determined by excitations not between delocalized well defined bands as in semiconductors but among localized spin-orbit coupled multiplet states \cite{Das2018}. While the overall dielectric functions show just a small variation across the phase transition, the condensation to an excitonic state of spin-orbit coupled states may influence the Raman tensor and the coherent nature of the oscillations. Therefore, further studies on the nature of the spin-phonon coupling in the $d^{4}$ Van Vleck state are desired to explain the coherent oscillations in Ca$_{2}$RuO$_{4}$. 

 The real-time observation of coherent oscillations in Ca$_2$RuO$_4$ manifests unique signatures of the spin-phonon coupling. It is surprising that the complicated interplay between spin and phonon could influence the birth of coherent phonons, which has neither been observed nor been expected before. As far as we know, there have been only two reports on abrupt oscillation-$phase$ shifts \cite{Schafer2010, Kamaraju2011}. As in the case of blue bronze, the observed $phase$ shift may take place concurrently in a few phonon modes. However, we could not recognize a noticeable anomaly in other phonon modes within our noise level. In a material that is transparent at the pump photon energy, such as Dy$_2$Ti$_2$O$_7$, the displacive mechanism can take effect only through multiphoton absorption, resulting in a small amplitude. Therefore, the oscillations should be susceptible to a phase transition that modifies the multiphoton absorption. The large oscillation-$phase$ shift in an opaque material like Ca$_{2}$RuO$_{4}$ requires a further theoretical study on the generation mechanism of the coherent phonons including the spin-phonon coupling in the Van Vleck type magnetic order of $d^{4}$ systems. Our results demonstrate that the $phase$-sensitive measurement of coherent oscillations offers a unique opportunity to investigate and control quantum phase transitions coupled to the lattice in complex materials.

\begin{acknowledgements}
We thank S.B. Chung for fruitful discussions. This work was supported by the Institute for Basic Science (IBS) in Korea (IBS-R009-D1). K.W.K. was supported by the Basic Science Research Program through the National Research Foundation of Korea (NRF) funded by the Ministry of Science, ICT and Future Planning (NRF-2015R1A2A1A10056200 and 2017R1A4A1015564). This work was supported by JSPS KAKENHI (Nos. JP15H05852, JP15K21717 and JP17H06136), JSPS Core-to-Core program. C.S acknowledges support of the JSPS International Research Fellowship (No. JP17F17027). E.-A.K. acknowledges support from the U.S. Department of Energy, Office of Basic Energy Sciences, Division of Materials Science and Engineering under Award DE-SC0010313 and from a Simons Fellowship in Theoretical Physics, Award \ce{\#}392182.
\end{acknowledgements}

\end{document}